\documentclass[conference]{IEEEtran}
\IEEEoverridecommandlockouts

\usepackage{booktabs}
\usepackage{changepage}
\usepackage{hyperref}
\usepackage{cite}
\usepackage{amsmath,amssymb,amsfonts}
\usepackage{algorithmic}
\usepackage{graphicx}
\usepackage{textcomp}
\usepackage{xcolor}
\def\BibTeX{{\rm B\kern-.05em{\sc i\kern-.025em b}\kern-.08em
    T\kern-.1667em\lower.7ex\hbox{E}\kern-.125emX}}

\begin{document}

\title{Low-Complexity Audio Embedding Extractors
\thanks{The computational results presented were achieved using the Linz Institute of Technology (LIT) AI Lab Cluster. The LIT AI Lab is supported by the Federal State of Upper Austria. GW's work is supported by the European Research Council (ERC) under the European Union's Horizon 2020 research and innovation programme, grant agreement No 101019375 (Whither Music?).}
}

\author{
\IEEEauthorblockN{Florian Schmid$^1$, Khaled Koutini$^2$, Gerhard Widmer$^{1,2}$}
\IEEEauthorblockA{$^1$\textit{Institute of Computational Perception}, $^2$\textit{LIT Artificial Intelligence Lab}  \\
\textit{Johannes Kepler University}, Linz, Austria\\
\{florian.schmid, khaled.koutini\}@jku.at}
}

\maketitle

\begin{abstract}
Solving tasks such as speaker recognition, music classification, or semantic audio event tagging with deep learning models typically requires computationally demanding networks. General-purpose audio embeddings (GPAEs) are dense representations of audio signals that allow lightweight, shallow classifiers to tackle various audio tasks. The idea is that a single complex feature extractor would extract dense GPAEs, while shallow MLPs can produce task-specific predictions. If the extracted dense representations are general enough to allow the simple downstream classifiers to generalize to a variety of tasks in the audio domain, a single costly forward pass suffices to solve multiple tasks in parallel. In this work, we try to reduce the cost of GPAE extractors to make them suitable for resource-constrained devices. We use efficient MobileNets trained on AudioSet using Knowledge Distillation from a Transformer ensemble as efficient GPAE extractors. We explore how to obtain high-quality GPAEs from the model, study how model complexity relates to the quality of extracted GPAEs, and conclude that low-complexity models can generate competitive GPAEs, paving the way for analyzing audio streams on edge devices w.r.t.~multiple audio classification and recognition tasks.

\end{abstract}


\section{Introduction}

Audio signals are high-dimensional and shallow representations, making them rarely useful for discriminative tasks without additional transformations or processing with complex models. Transforming raw audio signals into dense, low-dimensional audio embeddings allows a lightweight classifier to learn a task from limited amounts of labeled data~\cite{Turian21HEAR}. If the extracted audio embeddings are general, the raw audio signal must only be processed once, while task-specific downstream classifiers can produce the predictions for multiple tasks in parallel.

Historically, handcrafted low-dimensional representations were obtained by applying digital signal processing techniques~\cite{Liu1998DSPFeatureExtraction, Eyben2010openSMILE}
or audio signal transformations, such as calculating Mel Frequency Cepstral Coefficients~\cite{Logan2000MFCCs}. More recently, deep neural networks (DNNs) trained on large datasets have been able to extract more abstract, high-level representations~\cite{Hershey2017LargeScaleCNN, Cramer2019LLL, Oord2018ContrPredCod}. Architectures to extract dense audio representations include Convolutional Neural Networks (CNNs) processing 2D spectrograms~\cite{Hershey2017LargeScaleCNN, Cramer2019LLL, Niizumi2021ByolA}, 1D-CNNs operating directly on the waveform~\cite{Baevski2020wave2vec2, Oord2016wavenet} and Recurrent Neural Networks (RNNs) for modeling temporal dependencies~\cite{Mehri2017samplernn}. Recently, vision transformers~\cite{Dosovitskiy20Image, He2022MaeVit} have been ported to the audio domain~\cite{Gong21Ast, Koutini21Passt, niizumi22msm, Chong2022msp, Baade2022mae-ast} showing excellent audio classification and general-purpose audio extraction results. Models capable of extracting high-quality audio representations are typically trained on large datasets, such as ImageNet~\cite{Deng09ImageNet} for vision or AudioSet~\cite{audioset2017Gemmeke} for audio. Models are either trained in a supervised fashion~\cite{Koutini21Passt, Kong20PANNs} using labeled datasets or in a self-supervised way based on reconstruction~\cite{Baade2022mae-ast, niizumi22msm, Chong2022msp, Huang22Masked} and contrastive losses~\cite{Gong2022ssast, Niizumi2021ByolA, Saeed21contrastive}. In either regime, models tend to be complex to capture detailed feature representations.

To assess the quality of a general-purpose audio embedding extractor (GPAEE), benchmarks such as HEAR~\cite{Turian21HEAR} and HARES~\cite{Wang22hares} have been introduced. The GPAEE generates dense audio embeddings, while shallow classifiers are trained to perform task-specific predictions based on them. HEAR~\cite{Turian21HEAR} and HARES~\cite{Wang22hares} force the extracted embeddings to be universal by evaluating them on a variety of different audio tasks concerning speech, music or environmental sounds. Compared to fine-tuned models, generating predictions for multiple tasks requires only one costly feature extraction and several lightweight prediction steps. Reducing the computational demand of the GPAEE is an important step toward fitting this framework on resource-constrained devices. Prior work in this direction includes training and inference of self-supervised audio representation learning models on mobile devices~\cite{Tagliasacchi19audio_representation_mobile} and AemNet~\cite{Lopez-Meyer21EfficientAudioEmbeddings}, a model designed for efficient end-to-end audio embedding extraction. While the latter is the closest to our work, we use models with higher pre-training performance, test features extracted from different positions in the CNN, and evaluate on a much broader range of tasks.

In particular, we evaluate the performance of efficient MobileNets~\cite{Howard19MobileNetV3} trained on AudioSet~\cite{audioset2017Gemmeke} using Knowledge Distillation from a Transformer ensemble~\cite{Schmid22Efficient}\footnote{Pre-trained Models and Code are released on GitHub: \\ Pre-Trained Models: \href{https://github.com/fschmid56/EfficientAT}{https://github.com/fschmid56/EfficientAT} \\ HEAR evaluation: \href{https://github.com/fschmid56/EfficientAT\_HEAR}{https://github.com/fschmid56/EfficientAT\_HEAR}} as GPAEE on the HEAR benchmark~\cite{Turian21HEAR}. The contribution of this work is (1) to investigate how well-performing general-purpose audio representations can be obtained from a CNN, and (2) to analyze how the model complexity is related to the quality of extracted representations. As part of (2), we focus on low-complexity models and compare the parameter and computational efficiency of our proposed models to other single model GPAEEs.  

\section{Pre-trained MobileNets as General-purpose Audio Embedding Extractors}
\label{sec:pre-trained}
Efficient network design is a key enabler for deep learning on edge devices and has been well-studied in prior work~\cite{Howard17MobileNets, Sandler18MobileNetsV2, Howard19MobileNetV3, Tan19EfficientNet, Tan21EfficientNetV2}. Efficient vision architectures, such as EfficientNet~\cite{Tan19EfficientNet, Tan21EfficientNetV2} and MobileNet~\cite{Howard17MobileNets, Sandler18MobileNetsV2, Howard19MobileNetV3} have been ported to the audio domain and show an excellent performance-complexity trade-off~\cite{Gong21PSLA, Gong22CMKD, Schmid22Efficient}. In our experiments, we use MobileNetV3~\cite{Howard19MobileNetV3}, an architecture designed for the application on resource-constrained devices. The key building block of MobileNets is the mobile inverted bottleneck block~\cite{Sandler18MobileNetsV2}, a factorized design that is more computation- and memory-efficient than conventional convolutional layers. Squeeze-and-Excitation layers~\cite{Hu18Squeeze} are integrated into some blocks to recalibrate the filter responses and increase performance.

MobileNets pre-trained on AudioSet~\cite{audioset2017Gemmeke} using Knowledge Distillation from Transformers achieve state-of-the-art audio tagging performance, despite being much more efficient, in terms of memory, computation and model complexity than other models of similar performance~\cite{Schmid22Efficient}. 
We will scale the MobileNets by the width of the network, meaning that the number of layers stays constant, but the number of input and output channels per layer are multiplied by a factor $\alpha$. We abbreviate MobileNet as \textit{mn} and attach $\alpha$. In this sense, \textit{mn10} denotes the baseline model using $\alpha=1.0$ and consisting of 4.88M parameters. Setting $\alpha < 1$ produces models with reduced complexity while $\alpha > 1$ increases the complexity. Changing $\alpha$ modifies a model's computational and parameter complexity by roughly $\alpha^2$, allowing easy adaptation of the model's complexity for specific use cases~\cite{Howard17MobileNets}.

Table \ref{tab:mn10} depicts the network structure of \textit{mn10} consisting of an input convolution (in\_conv), 15 Blocks (B1-15) and a classification head, including a 1x1 convolution + global pooling (Clf 1), and two linear layers (Clf 2 and Clf 3) to predict the 527 classes of AudioSet~\cite{audioset2017Gemmeke}. Scaling the model width by $\alpha$ scales the size of extracted embeddings accordingly, i.e. \textit{mn20} doubles also the numbers \textit{\# out channels} and \textit{\# SE bottleneck} presented in Table \ref{tab:mn10}. In the following, we describe how we obtain dense representations from this model.

\begin{table}[]
\begin{tabular}{c|c|c|c}
\toprule
\textbf{Descriptor} & \textbf{\# out channels} & \textbf{\# SE bottleneck} & \textbf{stride} \\
\midrule
in\_conv            & 16                                                                & -                                                                    & 2               \\
B1                  & 16                                                                & -                                                                    & 1               \\
B2                  & 24                                                                & -                                                                    & 2               \\
B3                  & 24                                                                & -                                                                    & 1               \\
B4                  & 40                                                                & 24                                                                   & 2               \\
B5, 6               & 40                                                                & 32                                                                   & 1               \\
B7                  & 80                                                                & -                                                                    & 2               \\
B8, 9, 10           & 80                                                                & -                                                                    & 1               \\
B11                 & 112                                                               & 120                                                                  & 1               \\
B12                 & 112                                                               & 168                                                                  & 1               \\
B13                 & 160                                                               & 168                                                                  & 2               \\
B14, 15             & 160                                                               & 240                                                                  & 1               \\
Clf 1 (conv, avg. pool)  & 960                                                               & -                                                                    & 1               \\
Clf 2 (linear)      & 1280                                                              & -                                                                    & -               \\
Clf 3               & 527                                                               & -                                                                    & -     \\   
\bottomrule
\end{tabular}
\vspace{2 pt}
\caption{MobileNetV3~\cite{Howard19MobileNetV3} network structure and sizes using a width multiplier of 1.0 (\textit{mn10}). \# out channels denotes the number of channels as the output of blocks, \# SE bottleneck denotes the bottleneck size of the Squeeze-and-Excitation~\cite{Hu18Squeeze} layers.} 
\label{tab:mn10}
\vspace{-20 pt}
\end{table}

\subsection{High-Level Features}

High-level features are the most abstract representations. Corresponding to Table \ref{tab:mn10}, \textit{Clf 1} denotes the features resulting from global pooling, \textit{Clf 2} is the embedding from the penultimate linear layer and \textit{Clf 3} are the logits predicted for the 527 classes of AudioSet~\cite{audioset2017Gemmeke}. We denote the extracted embeddings as \textit{H\_Clf1} through \textit{H\_Clf3}. Very commonly, the feature representations \textit{H\_Clf1} or \textit{H\_Clf2} are used as fixed-size representations extracted from a CNN~\cite{Lopez-Meyer21EfficientAudioEmbeddings, Kong20PANNs}.

\subsection{Mid-Level Features}
\label{subsec:mid}

Mid-level features are extracted from the intermediate layers of the model. We compare two types of mid-level features: Squeeze-and-Excitation (SE) features extracted from the SE bottleneck layer, and block features extracted from the block output feature maps using global average pooling. We experimented with using max pooling instead of average pooling or the sum of both, but we found that using only average pooling yields the best results. The corresponding dimensionality of extracted embeddings for each block of \textit{mn10} is listed in Table \ref{tab:mn10} (\textit{\# out channels}, \textit{\# SE bottleneck}). We experimentally observed that more abstract representations obtained from higher-level blocks outperform lower-level representations on most of the tasks. We choose to concatenate three higher-level representations obtained from B11, B13 and B15 and one lower-level representation obtained from B5 for both block output and SE features. We ensured that removing any of the aforementioned blocks decreases overall performance on HEAR~\cite{Turian21HEAR}. We denote these two types of embeddings as \textit{M\_B} and \textit{M\_SE}.

\subsection{Low-Level Features}

We use mel spectrograms as input to our model. Mono audio is sampled at 32 kHz and STFT is computed using 25 ms windows and a hop size of 10 ms. Log mel spectrograms with 128 frequency bands are computed and serve as input to the models. Since global pooling removes time and pitch information from mid-level features, we add pitch information through low-level features by averaging the log mel spectrograms over time and denote this set of low-level features as \textit{L}.
Compared to pooling mid-level features only over time~\cite{Niizumi22multilayer}, which increases the embedding size by a factor that corresponds to the size of the frequency dimension in the feature map, our approach uses a fixed-size vector of 128 numbers, independent of the model's size.

\subsection{Scene and Timestamp Embeddings}

Common audio tasks require GPAEEs to generate embeddings for an entire audio clip (scene embeddings) or at regular intervals (timestamp embeddings)~\cite{Turian21HEAR}. To obtain scene embeddings we split audio clips into 10 seconds frames and average the resulting embeddings. For timestamp embeddings, we chunk the raw audio waveforms into overlapping windows of 160 ms with a hop size of 50 ms, similar to~\cite{Koutini22PaSSTHear}.

\section{Evaluation of Audio Representations}
\label{sec:evaluation}

A well-known testbed to assess the quality of extracted audio representations is the Holistic Evaluation of Audio Representations (HEAR) benchmark~\cite{Turian21HEAR} launched as a NeurIPS 2021 challenge. HEAR comprises 19 tasks with short and long time spans, covering different audio domains such as speech, music and environmental sounds. In an attempt to enforce universal audio representations, the range of downstream tasks is extremely broad, ranging from detecting the location of a gunshot to discriminating normal vs. queen-less beehives to classifying emotion in speech. We refer the reader to \cite{Turian21HEAR} for a detailed description of tasks and challenge models.

We use the HEAR-eval tool~\cite{Turian21HEAR} to evaluate all models and their extracted representations to be comparable to all submissions to the HEAR 2021 challenge. The evaluation follows two steps: firstly, embeddings for all tasks are generated using the GPAEE and, secondly, a task-specific shallow MLP is trained on the embeddings. 

\subsection{Evaluation Metrics}
\label{subsec:metrics}

The HEAR tasks use different evaluation metrics, such as Onset FMS, accuracy or mAP. To make the individual tasks comparable, we adopt the procedure in~\cite{Koutini21Passt} and normalize each score by the maximum score achieved by a model in the official HEAR 2021 challenge~\cite{Turian21HEAR}. The normalization allows to express the performance of each model on a task as a \textit{percentage of the best-performing challenge system}. We also adopt the grouping of tasks into \textit{speech}, \textit{music} and \textit{general} sounds presented in~\cite{Koutini21Passt}. 

Expressing the model performance in the benchmark as a single number is avoided in~\cite{Turian21HEAR} as it obscures nuances and details of model performances on individual tasks. Following this line, we present a detailed comparison between different single models in Section \ref{subsec:comp}. However, for studying individual performance factors of our models, we average the normalized scores across all tasks to derive a metric that can be interpreted as the \textit{average percentage of best-performing challenge systems}.

\section{Results}
\label{sec:results}

We first study which combinations of feature sets introduced in Section~\ref{sec:pre-trained} achieve the highest overall performance on HEAR~\cite{Turian21HEAR}. We then scale our models from 0.12 million parameters to 68 million parameters and test how the quality of extracted embeddings relates to the model complexity. Finally, we compare our proposed models to other single models evaluated on HEAR.

\subsection{Importance of Low-, Mid- and High-Level Features}
\label{subsec:importance}

In this section, we compare the performance of low-, \mbox{mid-,} high-level, and combined feature sets. We report all results based on \textit{mn10} (4.88M parameters) and the evaluation metric introduced in Section~\ref{subsec:metrics}.

\begin{table}[h]
\begin{center}
\begin{small}
\begin{tabular}{l|c|c|c|c||c}
\textbf{Feature Sets} &\textbf{\# dim} & \textbf{General} & \textbf{Music} & \textbf{Speech} & \textbf{All} \\ 
\midrule
L & 128 & 55.64 & \textbf{88.28} & 43.12 & 64.38 \\
M\_B & 472 & \textbf{93.01} & 74.66 & \textbf{76.52} & \textbf{81.91} \\
M\_SE & 560 & 87.07 & 70.77 & 74.14 & 77.66 \\
H\_Clf1 & 960 & 89.10 & 67.62 & 69.95 & 76.15 \\
H\_Clf2 & 1280 & 85.80 & 65.87 & 64.67 & 72.90 \\
H\_Clf3 & 527 & 73.50 & 55.86 & 42.31 & 58.79 \\
\midrule
M\_B+L & 600 & 87.75 & \textbf{91.10} & 77.23 & \textbf{86.22} \\
M\_B+M\_SE & 1032 & \textbf{91.65} & 73.07 & 77.32 & 81.03 \\
M\_B+H\_Clf1 & 1432 & 87.74 & 72.83 & \textbf{77.74} & 79.62 \\
M\_B+H\_Clf2 & 1752 & 89.75 & 72.20 & 73.61 & 79.04 \\
M\_B+H\_Clf3 & 999 & 76.88 & 61.92 & 53.90 & 65.32 \\
\bottomrule
\end{tabular}
\vspace{2pt}
\caption{Comparing different low-, mid-, and high-level feature sets based on the task categories General, Music, Speech and across all tasks.}
\label{tab:single}
\end{small}
\end{center}
\vspace{-20 pt}
\end{table}

Table \ref{tab:single} compares single feature sets in the first section and the concatenation of the best performing single feature set \textit{M\_B} with other feature sets in the second section. Regarding single feature sets, the logits \textit{H\_Clf3} containing information on the 527 AudioSet classes perform the worst, showing that these concepts are too specific to generalize well to a variety of downstream tasks. The more general mid-level features \textit{M\_B} and \textit{M\_SE} outperform all high-level features, indicating that high-level features are too specialized in the pre-training dataset domain. The low-level features \textit{L} show no good overall performance but achieve the highest performance in the \textit{Music} category, suggesting that pitch information is important but not available in higher-level features. Overall, the features \textit{M\_B} perform the best across all categories except for \textit{Music}. The second section shows that only the concatenation of the low-level features \textit{M\_B+L} brings an additional performance boost, as it adds the pitch information necessary to perform well in music-related tasks. Concatenating other feature sets to \textit{M\_B} does not improve overall performance, indicating that \textit{M\_B} already covers a large range of information extracted by the model from the raw audio signal.

\begin{figure*}[t!]
\centering
{\includegraphics[width=\textwidth]{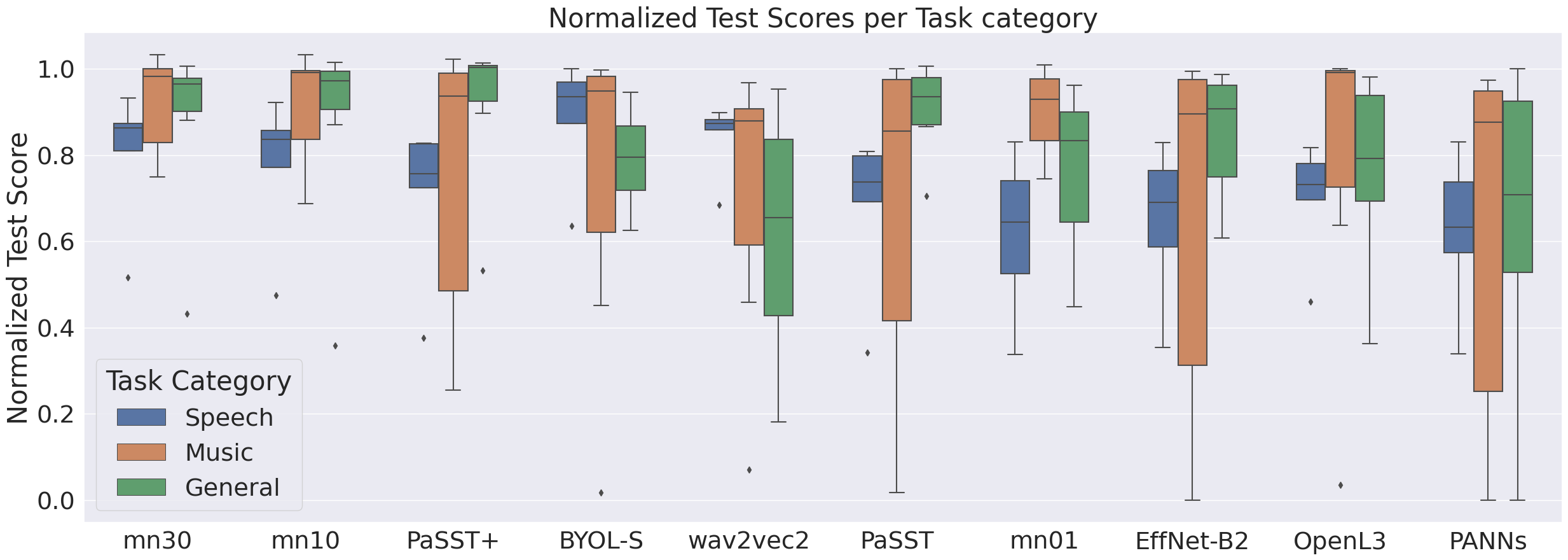}}
\caption{Comparing well-performing single models: PaSST and PaSST+~\cite{Koutini21Passt, Koutini22PaSSTHear} (supervised, Transformers), wav2vec2~\cite{Baevski2020wave2vec2} (self-supervised, Transformer),  BYOL-S~\cite{Elbanna22Byol-S} and OpenL3~\cite{Cramer2019LLL} (self-supervised, CNNs), and RedRice/Xiaomi EffNet-B2~\cite{Tan19EfficientNet} and PANNs CNN14~\cite{Kong20PANNs} (supervised, CNNs) to \textit{mn01}, \textit{mn10} and \textit{mn30}. We compare the distributions of normalized scores per task category between the models. The models are ordered according to the median value of the normalized scores across all tasks in descending order from left to right.}
\label{fig:category_analysis}
\vspace{-12 pt}
\end{figure*}

\begin{figure}[h]
\centering
{\includegraphics[width=\columnwidth]{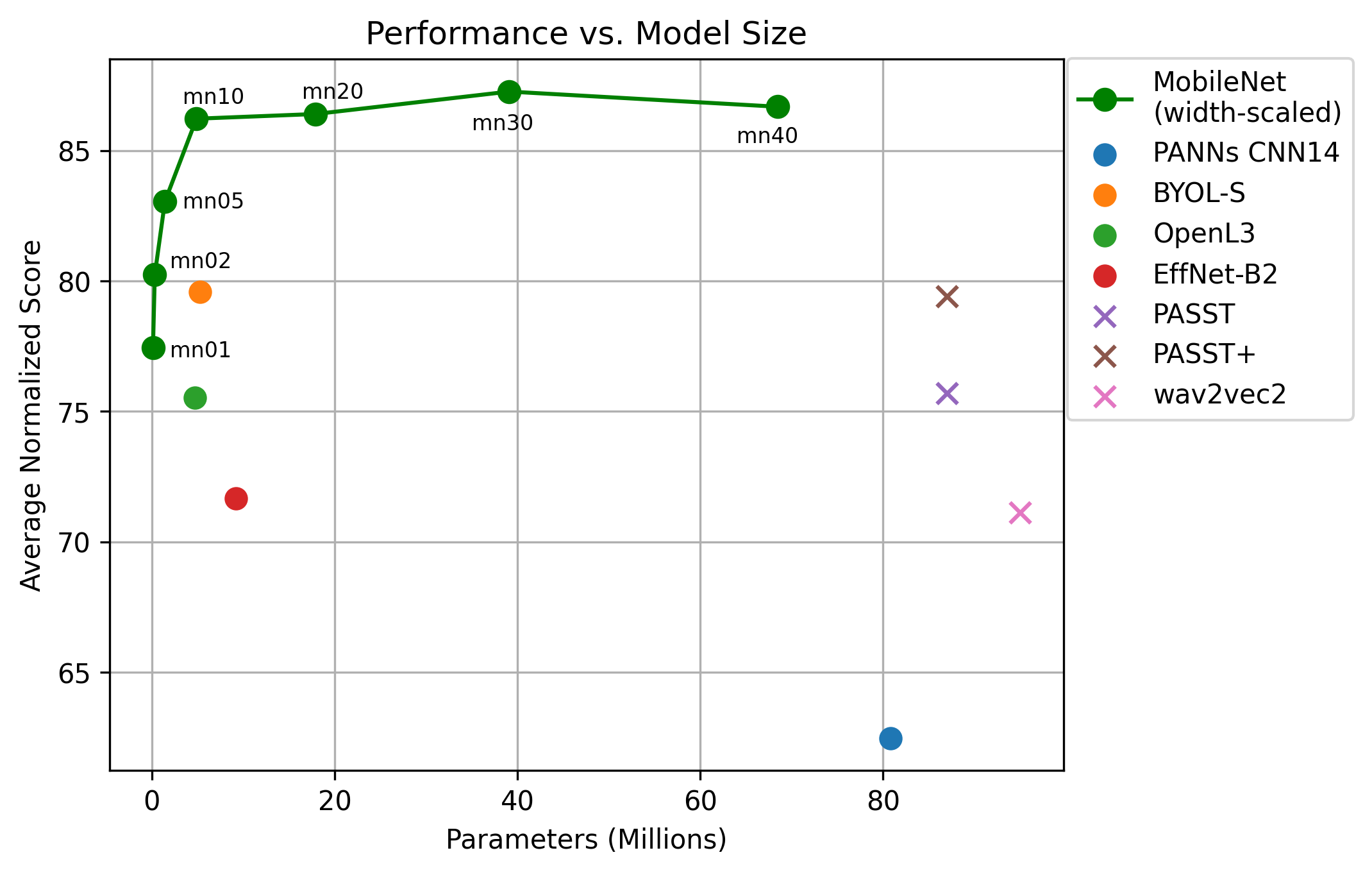}}
\caption{Comparing width-scaled ($\alpha \in \{0.1, 0.2, 0.5, 1.0, 2.0, 3.0, 4.0\}$) MobileNets to other well-performing single models with respect to the parameter-efficiency. Circles/Crosses denote CNNs/Transformers.}
\label{fig:model_sizes}
\vspace{-24pt}
\end{figure}

\subsection{Low-complexity Audio Embedding Extraction}
\label{sec:low-complexity}

In this section, we scale our models by adapting the width scaling factor $\alpha$ and compare the performance of models ranging from 0.12M (\textit{mn01}) to 68M parameters (\textit{mn40}). For all models, we use the best-performing feature set (\textit{M\_B+L}) found in Section \ref{subsec:importance}.

\textbf{Parameter Complexity}: Figure \ref{fig:model_sizes} compares the number of parameters against the average normalized scores of our width-scaled models ($\alpha \in \{0.1, 0.2, 0.5, 1.0, 2.0, 3.0, 4.0\}$) in comparison to well-performing single models submitted to the HEAR 21 challenge~\cite{Turian21HEAR}. In addition, we include PaSST+~\cite{Koutini22PaSSTHear}, an improved version of PaSST~\cite{Koutini21Passt} that uses a smaller hop size and concatenates mel features with two receptive fields of different sizes for timestamp embeddings. 

On the low-complexity end, our proposed models show an excellent parameter-performance trade-off. The smallest model \textit{mn01} performs worse than larger MobileNets and is outperformed by BYOL-S~\cite{Elbanna22Byol-S} (CNN trained in a self-supervised fashion) and PaSST+~\cite{Koutini22PaSSTHear} (Transformer trained on AudioSet~\cite{audioset2017Gemmeke} labels). However, with 120k parameters, it contains a fraction of the parameters of the other models and still provides audio embeddings of very competitive performance.

The performance of our models increases sharply until mn10 (4.88M parameters), with more complex models being only slightly better. mn30 reaches the highest performance, indicating that scaling up our models further is hitting a performance limit.

\textbf{Computational Complexity}: Complementary to the model size, the computational complexity at inference time specified in terms of multiply-accumulate (MAC) operations is an important factor for deploying models on resource-constrained devices. The computational demand of our models ranges from 20M (\textit{mn01}) over 540M (\textit{mn10}) up to 8B (\textit{mn40}) MACs per 10 seconds of processed audio signal. In comparison to other models (PaSST: 128B, CNN14: 20B, EffNet-B2: 900M, Byol-S: 780M), our pre-trained MobileNets are computationally lightweight. For example, \textit{mn01} could run real-time inference on an embedded processor supporting single-cycle MAC and operating in low MHz range, such as a Cortex-M4.

\subsection{Comparison to Single Models}
\label{subsec:comp}

Figure \ref{fig:category_analysis} compares \textit{mn01, mn10} and \textit{mn30} to other well-performing single models submitted to HEAR based on the normalized score distributions for the categories \textit{Speech}, \textit{Music} and \textit{General}. In the \textit{Speech} category, our models are outperformed by \textit{BYOL-S}~\cite{Elbanna22Byol-S} and \textit{wav2vec2}~\cite{Baevski2020wave2vec2}, two models that are specialized in speech tasks. However, \textit{mn10} and \textit{mn30} outperform all models not pre-trained on speech datasets. The \textit{Music} category is dominated by our models with \textit{mn01} exceeding the top challenge score on the task \textit{Beijing Opera Percussion} (recognize the type of percussion instrument) and \textit{mn30} setting new top scores for the tasks \textit{Mridingham Tonic} (classify tonics), \textit{Mridingham Stroke} (classify strokes) and \textit{GTZAN Genre} (classify genre). PaSST+ achieves the best results in the \textit{General} category, closely followed by our models. \textit{mn10} sets a new top score on the task \textit{Vocal Imitations} (retrieve sound using vocal imitation) and \textit{mn30} exceeds the top score on \textit{ESC-50} (environmental sound classification). Overall, \textit{mn30} and \textit{mn10} compare favorably against all other single models, being among the best models in each category. The tiny model \textit{mn01} performs comparable to EffNet-B2~\cite{Tan19EfficientNet}, OpenL3~\cite{Cramer2019LLL} and PANNs CNN14~\cite{Kong20PANNs}, and slightly worse than PaSST~\cite{Koutini21Passt}.

\section{Conclusion}
In this work, we used recently introduced highly efficient state-of-the-art audio tagging models~\cite{Schmid22Efficient} pre-trained on AudioSet~\cite{audioset2017Gemmeke} as low-complexity general-purpose audio embedding extractors. We tested which feature sets generalize best to a variety of downstream tasks and found that mid-level features perform best while adding low-level features brings in the necessary pitch information to master music-related tasks. Based on these findings, we varied the model complexity and showed that our scaled models are more parameter efficient and less computationally demanding than previously proposed models. We propose a tiny model \textit{mn01} that extracts audio embeddings of competitive quality for the application on edge devices. The larger \textit{mn10} is still very compact and compares favorably against other single models submitted to the HEAR challenge and, finally, the larger \textit{mn30} outperforms \textit{mn10} and beats the top HEAR challenge test scores on 4 tasks. 

\bibliographystyle{IEEEtran}
\bibliography{refs}

\begin{thebibliography}{10}
\providecommand{\url}[1]{#1}
\csname url@samestyle\endcsname
\providecommand{\newblock}{\relax}
\providecommand{\bibinfo}[2]{#2}
\providecommand{\BIBentrySTDinterwordspacing}{\spaceskip=0pt\relax}
\providecommand{\BIBentryALTinterwordstretchfactor}{4}
\providecommand{\BIBentryALTinterwordspacing}{\spaceskip=\fontdimen2\font plus
\BIBentryALTinterwordstretchfactor\fontdimen3\font minus
  \fontdimen4\font\relax}
\providecommand{\BIBforeignlanguage}[2]{{%
\expandafter\ifx\csname l@#1\endcsname\relax
\typeout{** WARNING: IEEEtran.bst: No hyphenation pattern has been}%
\typeout{** loaded for the language `#1'. Using the pattern for}%
\typeout{** the default language instead.}%
\else
\language=\csname l@#1\endcsname
\fi
#2}}
\providecommand{\BIBdecl}{\relax}
\BIBdecl

\bibitem{Turian21HEAR}
J.~Turian, J.~Shier, H.~R. Khan, B.~Raj, B.~W. Schuller, C.~J. Steinmetz,
  C.~Malloy, G.~Tzanetakis, G.~Velarde, K.~McNally, M.~Henry, N.~Pinto,
  C.~Noufi, C.~Clough, D.~Herremans, E.~Fonseca, J.~H. Engel, J.~Salamon,
  P.~Esling, P.~Manocha, S.~Watanabe, Z.~Jin, and Y.~Bisk, ``{HEAR:} holistic
  evaluation of audio representations,'' in \emph{NeurIPS 2021 Competitions and
  Demonstrations Track}.\hskip 1em plus 0.5em minus 0.4em\relax {PMLR}, 2021.

\bibitem{Liu1998DSPFeatureExtraction}
Z.~Liu, Y.~Wang, and T.~Chen, ``Audio feature extraction and analysis for scene
  segmentation and classification,'' \emph{J. {VLSI} Signal Process.}, 1998.

\bibitem{Eyben2010openSMILE}
F.~Eyben, M.~W{\"{o}}llmer, and B.~W. Schuller, ``Opensmile: the munich
  versatile and fast open-source audio feature extractor,'' in
  \emph{Proceedings of the 18th International Conference on Multimedia}.\hskip
  1em plus 0.5em minus 0.4em\relax {ACM}, 2010.

\bibitem{Logan2000MFCCs}
B.~Logan, ``Mel frequency cepstral coefficients for music modeling,'' in
  \emph{{ISMIR}, 1st International Symposium on Music Information Retrieval},
  2000.

\bibitem{Hershey2017LargeScaleCNN}
S.~Hershey, S.~Chaudhuri, D.~P.~W. Ellis, J.~F. Gemmeke, A.~Jansen, R.~C.
  Moore, M.~Plakal, D.~Platt, R.~A. Saurous, B.~Seybold, M.~Slaney, R.~J.
  Weiss, and K.~W. Wilson, ``{CNN} architectures for large-scale audio
  classification,'' in \emph{{IEEE} International Conference on Acoustics,
  Speech and Signal Processing, {ICASSP}}.\hskip 1em plus 0.5em minus
  0.4em\relax {IEEE}, 2017.

\bibitem{Cramer2019LLL}
J.~Cramer, H.~Wu, J.~Salamon, and J.~P. Bello, ``Look, listen, and learn more:
  Design choices for deep audio embeddings,'' in \emph{{IEEE} International
  Conference on Acoustics, Speech and Signal Processing, {ICASSP}}.\hskip 1em
  plus 0.5em minus 0.4em\relax {IEEE}, 2019.

\bibitem{Oord2018ContrPredCod}
A.~van~den Oord, Y.~Li, and O.~Vinyals, ``Representation learning with
  contrastive predictive coding,'' \emph{CoRR}, 2018.

\bibitem{Niizumi2021ByolA}
D.~Niizumi, D.~Takeuchi, Y.~Ohishi, N.~Harada, and K.~Kashino, ``{BYOL} for
  audio: Self-supervised learning for general-purpose audio representation,''
  in \emph{International Joint Conference on Neural Networks, {IJCNN}}.\hskip
  1em plus 0.5em minus 0.4em\relax {IEEE}, 2021.

\bibitem{Baevski2020wave2vec2}
A.~Baevski, Y.~Zhou, A.~Mohamed, and M.~Auli, ``wav2vec 2.0: {A} framework for
  self-supervised learning of speech representations,'' in \emph{Annual
  Conference on Neural Information Processing Systems, NeurIPS}, 2020.

\bibitem{Oord2016wavenet}
A.~van~den Oord, S.~Dieleman, H.~Zen, K.~Simonyan, O.~Vinyals, A.~Graves,
  N.~Kalchbrenner, A.~W. Senior, and K.~Kavukcuoglu, ``Wavenet: {A} generative
  model for raw audio,'' in \emph{The 9th {ISCA} Speech Synthesis
  Workshop}.\hskip 1em plus 0.5em minus 0.4em\relax {ISCA}, 2016.

\bibitem{Mehri2017samplernn}
S.~Mehri, K.~Kumar, I.~Gulrajani, R.~Kumar, S.~Jain, J.~Sotelo, A.~C.
  Courville, and Y.~Bengio, ``Samplernn: An unconditional end-to-end neural
  audio generation model,'' in \emph{5th International Conference on Learning
  Representations, {ICLR}}.\hskip 1em plus 0.5em minus 0.4em\relax
  OpenReview.net, 2017.

\bibitem{Dosovitskiy20Image}
A.~Dosovitskiy, L.~Beyer, A.~Kolesnikov, D.~Weissenborn, X.~Zhai,
  T.~Unterthiner, M.~Dehghani, M.~Minderer, G.~Heigold, S.~Gelly, J.~Uszkoreit,
  and N.~Houlsby, ``An image is worth 16x16 words: Transformers for image
  recognition at scale,'' in \emph{9th International Conference on Learning
  Representations, {ICLR}}.\hskip 1em plus 0.5em minus 0.4em\relax
  OpenReview.net, 2021.

\bibitem{He2022MaeVit}
K.~He, X.~Chen, S.~Xie, Y.~Li, P.~Doll{\'{a}}r, and R.~B. Girshick, ``Masked
  autoencoders are scalable vision learners,'' in \emph{{IEEE/CVF} Conference
  on Computer Vision and Pattern Recognition, {CVPR}}.\hskip 1em plus 0.5em
  minus 0.4em\relax {IEEE}, 2022.

\bibitem{Gong21Ast}
Y.~Gong, Y.~Chung, and J.~R. Glass, ``{AST:} audio spectrogram transformer,''
  in \emph{Interspeech, 22nd Annual Conference of the International Speech
  Communication Association}.\hskip 1em plus 0.5em minus 0.4em\relax {ISCA},
  2021.

\bibitem{Koutini21Passt}
K.~Koutini, J.~Schl{\"{u}}ter, H.~Eghbal{-}zadeh, and G.~Widmer, ``Efficient
  training of audio transformers with patchout,'' in \emph{Interspeech, 23rd
  Annual Conference of the International Speech Communication
  Association}.\hskip 1em plus 0.5em minus 0.4em\relax {ISCA}, 2022.

\bibitem{niizumi22msm}
D.~Niizumi, D.~Takeuchi, Y.~Ohishi, N.~Harada, and K.~Kashino, ``Masked
  spectrogram modeling using masked autoencoders for learning general-purpose
  audio representation,'' \emph{CoRR}, 2022.

\bibitem{Chong2022msp}
D.~Chong, H.~Wang, P.~Zhou, and Q.~Zeng, ``Masked spectrogram prediction for
  self-supervised audio pre-training,'' \emph{CoRR}, 2022.

\bibitem{Baade2022mae-ast}
A.~Baade, P.~Peng, and D.~Harwath, ``{MAE-AST:} masked autoencoding audio
  spectrogram transformer,'' in \emph{Interspeech, 23rd Annual Conference of
  the International Speech}.\hskip 1em plus 0.5em minus 0.4em\relax {ISCA},
  2022.

\bibitem{Deng09ImageNet}
J.~Deng, W.~Dong, R.~Socher, L.~Li, K.~Li, and L.~Fei{-}Fei, ``Imagenet: {A}
  large-scale hierarchical image database,'' in \emph{{IEEE} Computer Society
  Conference on Computer Vision and Pattern Recognition, {CVPR}}.\hskip 1em
  plus 0.5em minus 0.4em\relax {IEEE} Computer Society, 2009.

\bibitem{audioset2017Gemmeke}
J.~F. Gemmeke, D.~P.~W. Ellis, D.~Freedman, A.~Jansen, W.~Lawrence, R.~C.
  Moore, M.~Plakal, and M.~Ritter, ``Audio set: An ontology and human-labeled
  dataset for audio events,'' in \emph{{IEEE} International Conference on
  Acoustics, Speech and Signal Processing, {ICASSP}}.\hskip 1em plus 0.5em
  minus 0.4em\relax {IEEE}, 2017.

\bibitem{Kong20PANNs}
Q.~Kong, Y.~Cao, T.~Iqbal, Y.~Wang, W.~Wang, and M.~D. Plumbley, ``Panns:
  Large-scale pretrained audio neural networks for audio pattern recognition,''
  \emph{{IEEE} {ACM} Trans. Audio Speech Lang. Process.}, 2020.

\bibitem{Huang22Masked}
P.~Huang, H.~Xu, J.~Li, A.~Baevski, M.~Auli, W.~Galuba, F.~Metze, and
  C.~Feichtenhofer, ``Masked autoencoders that listen,'' in \emph{Annual
  Conference on Neural Information Processing Systems, NeurIPS}, 2022.

\bibitem{Gong2022ssast}
Y.~Gong, C.~Lai, Y.~Chung, and J.~R. Glass, ``{SSAST:} self-supervised audio
  spectrogram transformer,'' in \emph{Thirty-Sixth {AAAI} Conference on
  Artificial Intelligence}.\hskip 1em plus 0.5em minus 0.4em\relax {AAAI}
  Press, 2022.

\bibitem{Saeed21contrastive}
A.~Saeed, D.~Grangier, and N.~Zeghidour, ``Contrastive learning of
  general-purpose audio representations,'' in \emph{{IEEE} International
  Conference on Acoustics, Speech and Signal Processing, {ICASSP}}.\hskip 1em
  plus 0.5em minus 0.4em\relax {IEEE}, 2021.

\bibitem{Wang22hares}
L.~Wang, P.~Luc, Y.~Wu, A.~Recasens, L.~Smaira, A.~Brock, A.~Jaegle,
  J.~Alayrac, S.~Dieleman, J.~Carreira, and A.~van~den Oord, ``Towards learning
  universal audio representations,'' in \emph{{IEEE} International Conference
  on Acoustics, Speech and Signal Processing, {ICASSP}}.\hskip 1em plus 0.5em
  minus 0.4em\relax {IEEE}, 2022.

\bibitem{Tagliasacchi19audio_representation_mobile}
M.~Tagliasacchi, B.~Gfeller, F.~de~Chaumont~Quitry, and D.~Roblek,
  ``Self-supervised audio representation learning for mobile devices,''
  \emph{CoRR}, 2019.

\bibitem{Lopez-Meyer21EfficientAudioEmbeddings}
P.~Lopez{-}Meyer, J.~A. del Hoyo~Ontiveros, H.~Lu, and G.~Stemmer, ``Efficient
  end-to-end audio embeddings generation for audio classification on target
  applications,'' in \emph{{IEEE} International Conference on Acoustics, Speech
  and Signal Processing, {ICASSP}}.\hskip 1em plus 0.5em minus 0.4em\relax
  {IEEE}, 2021.

\bibitem{Howard19MobileNetV3}
A.~Howard, R.~Pang, H.~Adam, Q.~V. Le, M.~Sandler, B.~Chen, W.~Wang, L.~Chen,
  M.~Tan, G.~Chu, V.~Vasudevan, and Y.~Zhu, ``Searching for mobilenetv3,'' in
  \emph{{IEEE/CVF} International Conference on Computer Vision, {ICCV}}.\hskip
  1em plus 0.5em minus 0.4em\relax {IEEE}, 2019.

\bibitem{Schmid22Efficient}
F.~Schmid, K.~Koutini, and G.~Widmer, ``Efficient large-scale audio tagging via
  transformer-to-cnn knowledge distillation,'' in \emph{{IEEE} International
  Conference on Acoustics, Speech and Signal Processing, {ICASSP}}.\hskip 1em
  plus 0.5em minus 0.4em\relax {IEEE}, 2023.

\bibitem{Howard17MobileNets}
A.~G. Howard, M.~Zhu, B.~Chen, D.~Kalenichenko, W.~Wang, T.~Weyand,
  M.~Andreetto, and H.~Adam, ``Mobilenets: Efficient convolutional neural
  networks for mobile vision applications,'' \emph{CoRR}, 2017.

\bibitem{Sandler18MobileNetsV2}
M.~Sandler, A.~G. Howard, M.~Zhu, A.~Zhmoginov, and L.~Chen, ``Mobilenetv2:
  Inverted residuals and linear bottlenecks,'' in \emph{{IEEE} Conference on
  Computer Vision and Pattern Recognition, {CVPR}}.\hskip 1em plus 0.5em minus
  0.4em\relax Computer Vision Foundation / {IEEE} Computer Society, 2018.

\bibitem{Tan19EfficientNet}
M.~Tan and Q.~V. Le, ``Efficientnet: Rethinking model scaling for convolutional
  neural networks,'' in \emph{Proceedings of the 36th International Conference
  on Machine Learning, {ICML}}.\hskip 1em plus 0.5em minus 0.4em\relax {PMLR},
  2019.

\bibitem{Tan21EfficientNetV2}
------, ``Efficientnetv2: Smaller models and faster training,'' in
  \emph{Proceedings of the 38th International Conference on Machine Learning,
  {ICML}}.\hskip 1em plus 0.5em minus 0.4em\relax {PMLR}, 2021.

\bibitem{Gong21PSLA}
Y.~Gong, Y.~Chung, and J.~R. Glass, ``{PSLA:} improving audio tagging with
  pretraining, sampling, labeling, and aggregation,'' \emph{{IEEE} {ACM} Trans.
  Audio Speech Lang. Process.}, 2021.

\bibitem{Gong22CMKD}
Y.~Gong, S.~Khurana, A.~Rouditchenko, and J.~R. Glass, ``{CMKD:}
  cnn/transformer-based cross-model knowledge distillation for audio
  classification,'' \emph{CoRR}, 2022.

\bibitem{Hu18Squeeze}
J.~Hu, L.~Shen, and G.~Sun, ``Squeeze-and-excitation networks,'' in
  \emph{{IEEE} Conference on Computer Vision and Pattern Recognition,
  {CVPR}}.\hskip 1em plus 0.5em minus 0.4em\relax Computer Vision Foundation /
  {IEEE} Computer Society, 2018.

\bibitem{Niizumi22multilayer}
D.~Niizumi, D.~Takeuchi, Y.~Ohishi, N.~Harada, and K.~Kashino, ``Composing
  general audio representation by fusing multilayer features of a pre-trained
  model,'' in \emph{30th European Signal Processing Conference,
  {EUSIPCO}}.\hskip 1em plus 0.5em minus 0.4em\relax {IEEE}, 2022.

\bibitem{Koutini22PaSSTHear}
K.~Koutini, S.~Masoudian, F.~Schmid, H.~Eghbal{-}zadeh, J.~Schl{\"{u}}ter, and
  G.~Widmer, ``Learning general audio representations with large-scale training
  of patchout audio transformers,'' in \emph{HEAR}.\hskip 1em plus 0.5em minus
  0.4em\relax {PMLR}, 2023.

\bibitem{Elbanna22Byol-S}
G.~Elbanna, N.~Scheidwasser{-}Clow, M.~Kegler, P.~Beckmann, K.~E. Hajal, and
  M.~Cernak, ``{BYOL-S:} learning self-supervised speech representations by
  bootstrapping,'' in \emph{HEAR}.\hskip 1em plus 0.5em minus 0.4em\relax
  {PMLR}, 2023.

\end{thebibliography}

\end{document}